\pdfoutput=1
\documentclass[reprint,
amsmath,amssymb,
]{revtex4-2}

\usepackage{dcolumn}% Align table columns on decimal point
\usepackage[usenames]{color}
\usepackage{amssymb}
\usepackage{amsmath}
\usepackage{commath}
\usepackage{graphicx,bm}
\usepackage{verbatim}
\usepackage{float}

\begin{document}
\title{Dynamics of dissipative topological defects in coupled phase oscillators}
\author{Simon Mahler}\email{Corresponding author: sim.mahler@gmail.com}
\affiliation{Department of Physics of Complex Systems, Weizmann Institute of Science, Rehovot 7610001, Israel}
\author{Vishwa Pal}
\affiliation{Department of Physics, Indian Institute of Technology Ropar, Rupnagar 140001, Punjab, India}
\author{Chene Tradonsky}
\author{Ronen Chriki}
\author{Asher A. Friesem}
\author{Nir Davidson}
\affiliation{Department of Physics of Complex Systems, Weizmann Institute of Science, Rehovot 7610001, Israel}
\begin{abstract}
The dynamics of dissipative topological defects in a system of coupled phase oscillators, arranged in one and two-dimensional arrays, is numerically investigated using the Kuramoto model. After an initial rapid decay of the number of topological defects, due to vortex$-$anti-vortex annihilation, we identify a long-time (quasi) steady state where the number of defects is nearly constant. We find that the number of topological defects at long times is significantly smaller when the coupling between the oscillators is increased at a finite rate rather than suddenly turned on. Moreover, the number of topological defects scales with the coupling rate, analogous to the cooling rate in Kibble-Zurek mechanism (KZM). Similar to the KZM, the dynamics of topological defects is governed by two competing time scales: the dissipation rate and the coupling rate. Reducing the number of topological defects improves the long time coherence and order parameter of the system and enhances its probability to reach a global minimal loss state that can be mapped to the ground state of a classical XY spin Hamiltonian. 
\end{abstract}
\maketitle
%\tableofcontents
\section{Introduction}
Topological defects occur in many fields including atomic and molecular physics, condensed matter, cosmology, cold atoms, fluids mechanics, spin systems and optics \cite{art1,art2,art3,artau,Giomi,Pyka2013}. They appear across a phase transition and limit the order of systems. Their origin and behavior were first characterized by the Kibble-Zurek mechanism (KZM) as a system quenched across a phase transition into an ordered state via competing time scales \cite{art4,art5,art6}. In the KZM, the number of topological defects obeys a power law as a function of the quench rate. This power law was related to the critical exponent associated to the correlation length and the relaxation time of the system in a universal manner \cite{art5,ZUREK1996177}.

The KZM was investigated in diverse environments: atomic gases \cite{Navon167,art8,art9}, condensed matter systems \cite{art6,art11} and nonlinear optics \cite{art10}. Observing the KZM experimentally is subject to strong limitations, the cooling rate must be slow enough to keep the system homogeneous but not too slow to prevent finite size effect \cite{Navon167}. In particular for systems cooled from the outside this implies limitations into the interesting regime of fast cooling, where the system fails to maintain cooling uniformity \cite{art8,art11,art12,art13,Navon167}.

Recently, dissipative topological defects were numerically and experimentally investigated using a one-dimensional ring array of phased-locked lasers, where their formation was shown to be related to the KZM in a universal manner by two competing time scales \cite{vishwa}. The ratio between these two time scales depends on the system parameters, with which it is possible to enable the system to dissipate to a fully ordered, defect-free state that can be exploited for solving computational problems in various fields \cite{Marandi2014}. The dissipative topological defects were investigated with a fixed coupling between the lasers but intensity fluctuations acted as an effective heat bath with controllable cooling rate.  The lasers platform is essentially a system of coupled oscillators where the formation of topological defects occurs without external cooling. They reach a stable phase-locked state due to a dissipative coupling that minimizes the loss of the system, and can be mapped to the ground state of a classical XY spin Hamiltonian \cite{Marandi2014,Eckhouse,eitan}. With a complex landscape of many local minima, the system does not reach the ground state solution and remains stuck in local minima. For the ring geometry, these local minima have a nonzero topological charge (winding number) and were defined as dissipative topological defects \cite{vishwa,Palopticsexpress,distrib}. Under the assumption of constant field amplitudes, such coupled lasers are well approximated as Kuramoto phase oscillators \cite{Acebron05,eitan}. 

In this paper, we investigated the dynamics of topological defects in one and two-dimensional arrays of coupled oscillators, where we varied the coupling strenght in time to correspond to temperature variation in the KZM. In our investigation, we used the Kuramoto model \cite{kuramotobook}, where we assumed that all oscillators have the same amplitude and interact with nearest neighbors only. We swept the coupling in time at different rates to observe the dynamics of dissipative topological defects, and then established their connections with KZM where uncoupled oscillators are equivalent to the infinite temperature state and strongly coupled phase-locked oscillators are equivalent to a zero temperature state. In addition to the variation of the coupling, we showed that introducing disorder at the initial state leads to a more ordered stable state.
\section{Array of coupled oscillators and the Kuramoto model}
The Kuramoto model describes the temporal evolution of the phase of coupled oscillators \cite{kuramotobook,kuramotostrog}. It was successfully applied in many areas such as neural networks, complex systems, chemical and biological oscillators \cite{kuramotobook,Acebron05} and recently in coupled lasers for observing topological defects \cite{vishwa,Flovik2016}. The model is valid when all oscillators are nearly identical and the coupling between them is weak.

Our oscillators are coupled to their nearest neighbors ($NN$) only \cite{vishwa}. Their phases evolve in time as a function of the detuning and the coupling strength between the oscillators:

\begin{equation}
\frac{d\phi_{m}}{d(t/\tau_{p})}=~\Omega_{m}\tau_{p}+\sum_{n=(m)_{NN}}K_{mn}~\sin(\phi_{n}-\phi_{m}),
\label{eq:1_kuramoto_model}
\end{equation} 
where $\phi_{m}$ is the phase of the $m^{th}$ oscillator, $\Omega_{m}$ is the frequency detuning from a common frequency $\Omega_{0}=0$, $\tau_{p}$ is the lifetime between two iterations, $K_{mn}$ is the coupling between oscillators $m$ and $n$ and the sum occurs only on the nearest neighbors of the $m^{th}$ oscillator. 

Equation~(\ref{eq:1_kuramoto_model}) is solved using the fourth-order Runge-Kutta method. The time is expressed in unit of iterations $T=t/\tau_{p}$ and the initial phases of the oscillators were chosen randomly between $\left[-\pi ~\mathrm{to} ~\pi\right]$. Using Eq.~(\ref{eq:1_kuramoto_model}), we determined the dynamics of dissipative topological defects in a square and ring array of coupled phase oscillators, we show the relation between time dependent coupling strength and KZM, and we introduce disorder into the system. The results are presented in section III to VI.  
\section{Dissipative Topological Defects in a 2-D square array}
\begin{figure}[!ht]
\centering
\includegraphics[width=0.49\textwidth]{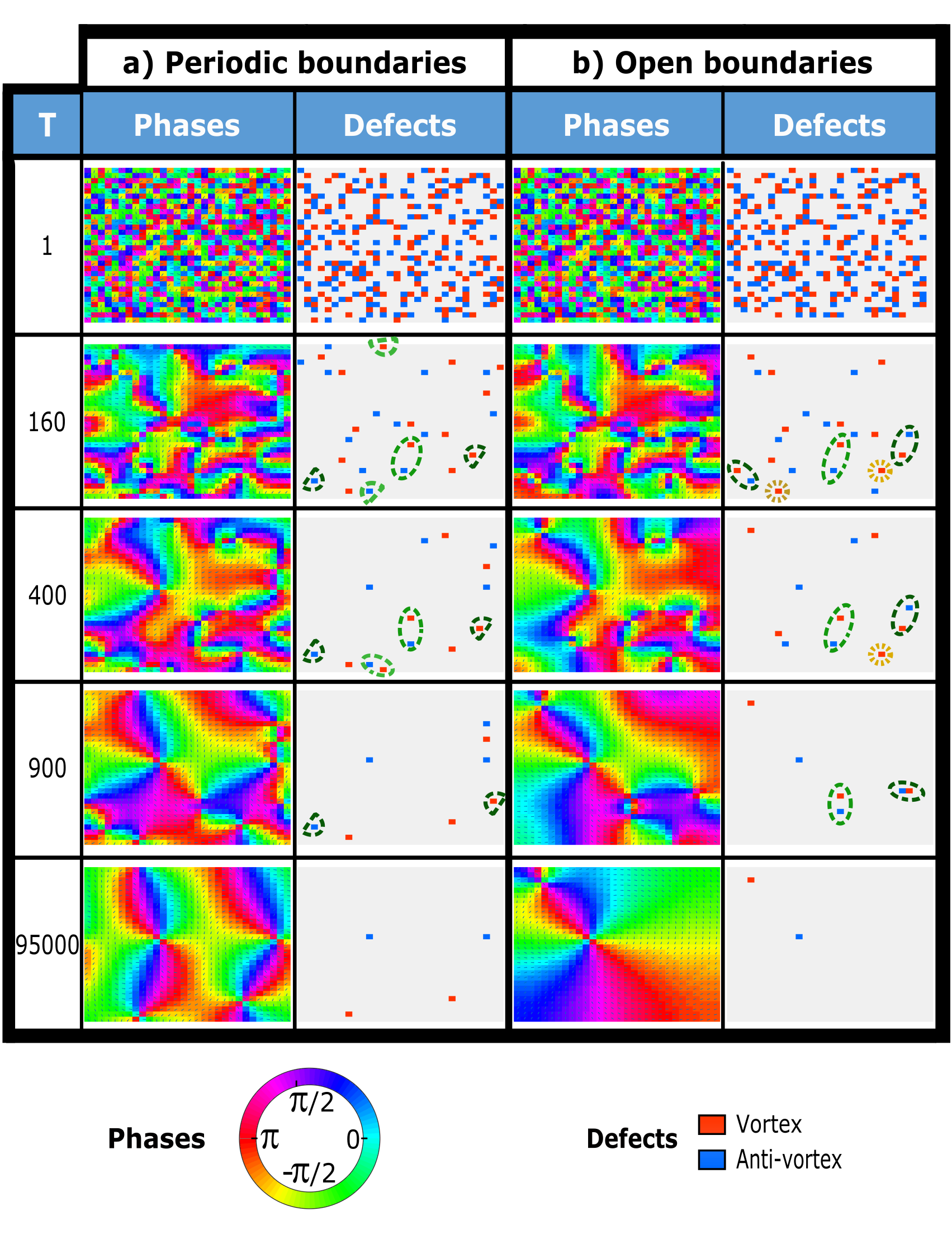}
\caption{Dynamics of topological defects in a square array of $900$ ($30\times30$) coupled oscillators with a) periodic and b) open boundary conditions. The second and fourth columns show the evolution in time of the phases and the third and fifth columns show the evolution of the topological defects. Initially ($T=1$), the number of topological defects is large, but as the iterations increase the number of topological defects decrease. The green dashed circles surround a pair of topological defects in which vortex and anti-vortex attract each other and eventually are annihilated. In the open boundary case, the yellow circles surround topological defects which eventually dissipate into the boundaries. In the simulation, the initial phases of the oscillators were uniformly distributed in the range $\left[-\pi ~\mathrm{to} ~\pi\right]$, $\Omega$=0, and $K=0.1$.}
\label{fig:1_eg_anhil}
\end{figure}

We investigated the dynamics of dissipative topological defects in a square array of $M=900$ oscillators, using Eq.~(\ref{eq:1_kuramoto_model}) to calculate the phases evolution. A topological defect is defined as either a vortex or an anti-vortex, identified by using a defect number $D$, where $D=1$ corresponds to a vortex and $D=-1$ to an anti-vortex (see Appendix~\ref{sec:Appendix_A}). We assumed that there were no detuning between the oscillators ($\Omega_{m}=0$), and that the coupling between the oscillators was kept constant $K_{mn}=0.1$. The results are presented in Fig.~\ref{fig:1_eg_anhil}.

Figure~\ref{fig:1_eg_anhil}(a) shows, for periodic boundary conditions, the simulated time evolution (iteration unit $T$) of the phases of the oscillators (second column) and defects (third column). As evident, the phases become smoother with time due to the dissipative dynamics and the number of phase singularities which connect the smooth regions decrease. The dissipative dynamics involve vortex$-$vortex, anti-vortex$-$anti-vortex, and vortex$-$anti-vortex interactions, where same charges repel each other, and opposite charges attract each other. Accordingly, vortex$-$anti-vortex attraction forms coupled pairs (surrounded by the green dashed circles in the third column of Fig.~\ref{fig:1_eg_anhil} that eventually annihilate. This annihilation process follows a well-defined potential, with logarithmic scaling \cite{hamerly}. After a long time ($T>95000$), few topological defects remain and the large distance between them critically slows down the anihilation process, leading to a quasi steady-state (stable state). The average distance between the topological defects can be then related to a correlation length thereby indicating the coherence of the system \citep{Navon167}. As the number of topological defects decrease, the distance between a pair and the correlation length of the system increase. Figure~\ref{fig:1_eg_anhil}(b) shows the simulated results for open boundary conditions. The same dissipative dynamics as for the periodic boundary conditions is observed. However, some topological defects dissipate into the boundaries. These are shown by the yellow star circles in the last column of Fig.~\ref{fig:1_eg_anhil}. As the system size is larger, the difference between periodic and open boundary conditions becomes smaller. In the thermodynamic limit (and also for a square array of $40000$ oscillators that we study below), there is no difference between the periodic and open boundary conditions.

\begin{figure}
\centering
\includegraphics[width=0.49\textwidth]{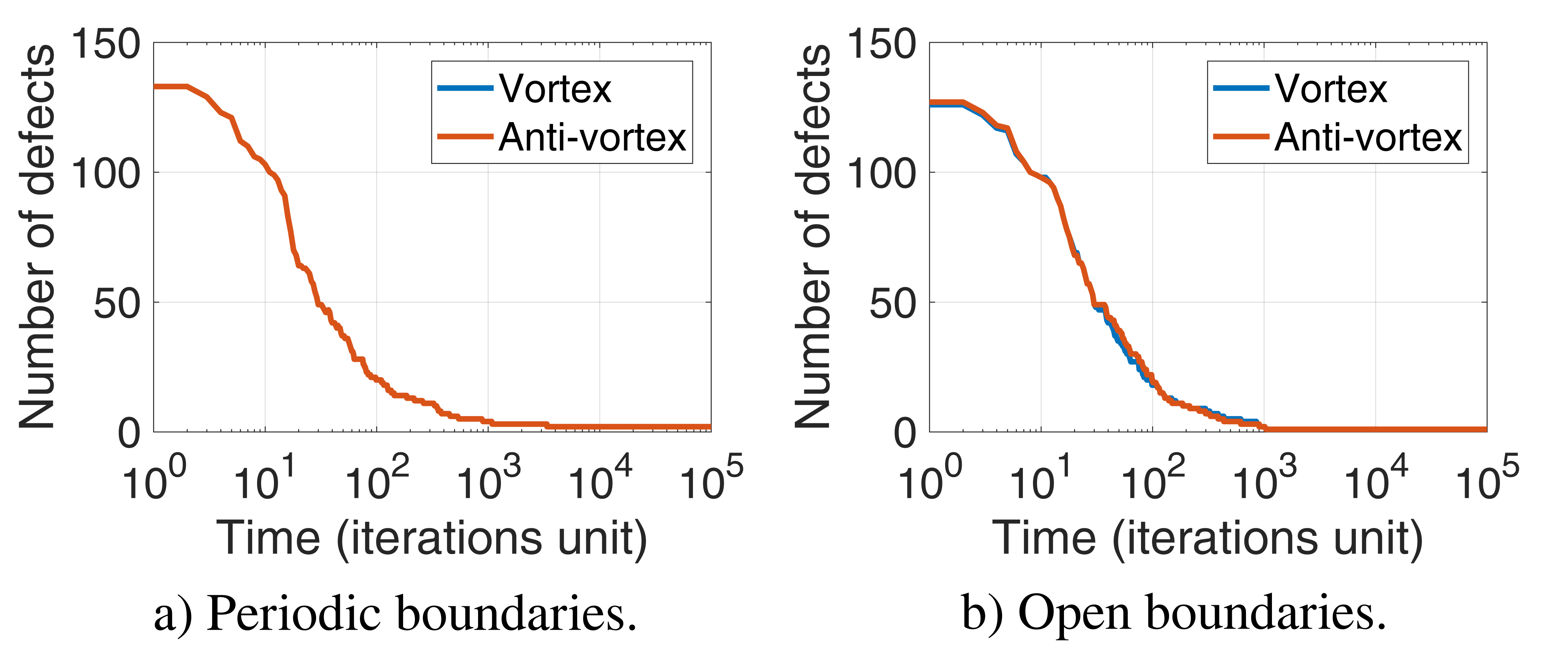}
\caption{The number of topological defects as a function of time for the square array shown in Fig. 1, with periodic and open boundaries. The number of topological defects decreases rapidly and later slowly, which indicates the existence of two time scales. Note the logarithmic scale along the horizontal axis.}
\label{fig:2_eg_anhil_number_def}
\end{figure}

Figure~\ref{fig:2_eg_anhil_number_def} shows a quantitative time evolution of the number of topological defects for both periodic and open boundary conditions. As evident, the number of topological defects is large in the initial state ($T=1$), and decreases in time due to the dissipative dynamics, until only few topological defects remain in the stable state ($T=95000$). For the periodic boundary conditions, Fig.~\ref{fig:2_eg_anhil_number_def}(a), both the number of vortices and anti-vortices curves overlap in time, indicating that they can only dissipate by annihilating each other. For the open boundary conditions, Fig.~\ref{fig:2_eg_anhil_number_def}(b), the number of vortices differs slightly from the number of anti-vortices indicating that topological defects can also dissipate into the boundaries.

To quantify the ordering in the system, we calculated the order parameter \cite{kuramotobook,kuramotostrog}:

\begin{equation}
r=\frac{1}{M}\left|\sum_{m=1}^{M}e^{i\phi_{m}}\right|,
\label{eq:2_order_param} 
\end{equation} 
where the sum occurs over all $M$ oscillators. Equation~(\ref{eq:2_order_param}) represents the analog of thermal averaging for a classical spin-system \cite{eitan}. The order parameter $r$ is in the range $\left[0 ~\mathrm{to} ~1\right]$. When $r=1$, all the oscillators have the same phase, i.e. a fully ordered state. Topological defect reduces the order parameter of the system. 

\begin{figure}[htb]
\centering
\includegraphics[width=0.49\textwidth]{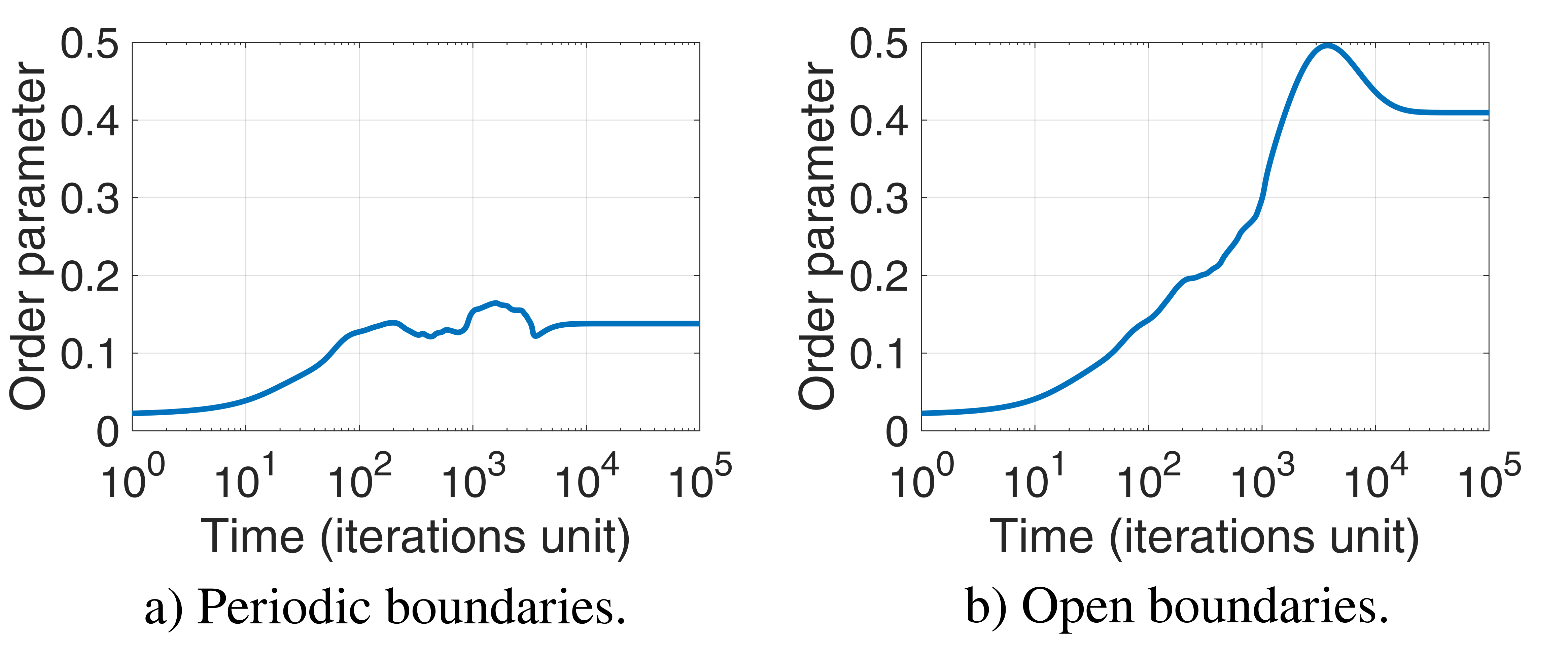}
\caption{The evolution of the order parameter as a function of time for the square array shown in Fig.~\ref{fig:1_eg_anhil}, with periodic and open boundaries. The order parameter increases with time, corresponding to the decrease in time of the number of topological defects.}
\label{fig:3_eg_anhil_order_param}
\end{figure}

Using Eq.~(\ref{eq:2_order_param}), we calculated the evolution in time of the order parameter in the square array of coupled oscillators shown in Fig. 1 with periodic and open boundary conditions. The results are presented in Fig.~\ref{fig:3_eg_anhil_order_param}. As shown, the order parameter increases from a low value (initial state) to a finite value (stable state). The higher stable state order parameter for the open boundaries ($0.4$ versus $0.15$) is consistent with the smaller number of topological defect  in Fig.~\ref{fig:2_eg_anhil_number_def}, due to the boundary dissipation. 

\section{Time-dependent coupling strength}
Next, we introduced time varying coupling (coupling rate) as an analogue of cooling rate in KZM \cite{Takeda2018}. For non-zero detuning, below a certain coupling strength the oscillators remain unsynchronized \cite{Nixon21, Fabiny, PhysRevLett651701}, in a disorder state. For coupling higher than a characteristic critical coupling, the oscillators phase-lock, to form an ordered state. Thus, varying coupling with time across the transition from disordered state to an ordered state, provides a more direct analogy to KZM \cite{art6}. Time varying coupling can be given as:
\begin{equation}
K(T)=
\begin{cases}
CT, & \text{if}\ CT\leq K_{max} \\
K_{max}, & \text{otherwise} 
\end{cases}
\label{eq:3_ramping_K}
\end{equation}
where $C$ is the quench rate, $K_{max}=2.5$ is  the maximal coupling (arbitrarily chosen). An infinite quench rate corresponds to a sudden jump in the coupling strength.

Using Eq.(\ref{eq:1_kuramoto_model}), we investigated the dynamics of dissipative topological defects and KZM for a square array of $M=40000$ oscillators with nearest neighbors coupling. The detuning between the oscillators was randomly chosen in the range $[0 ~\mathrm{to} ~\Omega_{max}\approx\tau_{p}^{-1}\pi/8]$ and the coupling was increased in time according to Eq.~(\ref{eq:3_ramping_K}). For this large system size, periodic and open boundary conditions give the same results. 

\begin{figure}[htb]
\centering
\includegraphics[width=0.49\textwidth]{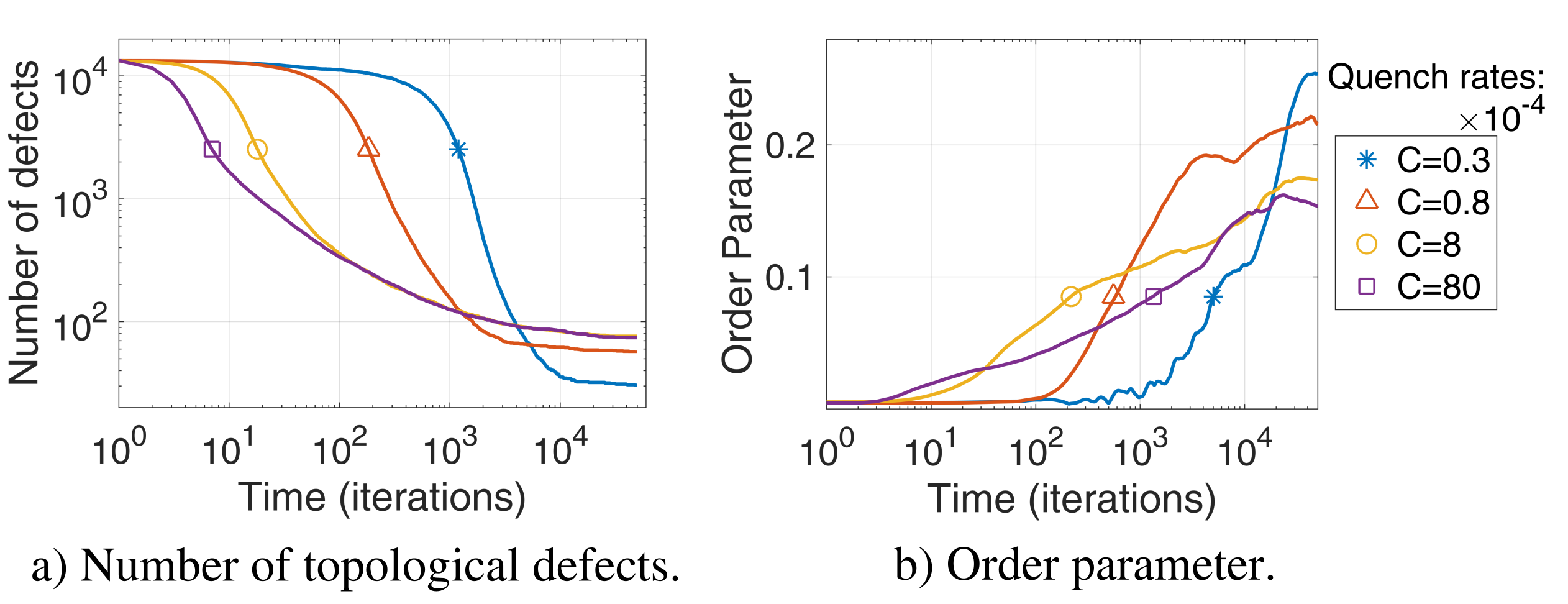}
\caption{The defects dynamics in a square array of $40000$ ($200\times200$) oscillators, with periodic boundaries, for four different quench rates of the coupling. a) The number of topological defects as a function of time. b) Corresponding evolution in time of the order parameter. The results were obtained by averaging over six different realizations. For each realization, the initial conditions of phases were chosen in the range of $\left[-\pi ~\mathrm{to} ~\pi\right]$ and the detuning in the range of $[0 ~\mathrm{to} ~\Omega_{max}\approx\tau_{p}^{-1}\pi/8]$.}
\label{fig:4_effect_eg_1}
\end{figure}

The results for the periodic boundary conditions are presented in Fig.\ref{fig:4_effect_eg_1}. Figure ~\ref{fig:4_effect_eg_1}(a) shows the number of topological defects as a function of time. At very high quench rate $C=80\times 10^{-4}$, the dynamics of the number of topological defects is similar to those of constant coupling  (Fig.~\ref{fig:2_eg_anhil_number_def}). Such high quench rate corresponds to an early sudden jump of the coupling (see Fig.~\ref{fig:S1_ramping_K_O}(a) of Appendix~\ref{sec:Appendix_A}). As the quench rate is lowered, the number of topological defects starts to decrease later but eventually reaches a lower value at the stable state. This long time crossing is also evident in the corresponding time evolution of order parameter, as shown in Fig.~\ref{fig:4_effect_eg_1}(b). 

In the stable state, the observation of increased topological defects with the increase in coupling rate is analogous to the KZM, where there are more defects at higher cooling rate \cite{art6}. We verified these observations for different system size (number of oscillators), and found a similar behaviour (see Appendix~\ref{sec:Appendix_B}), where the initial number of topological defects and the number of topological defects at the stable state increase with the system size (see Appendix~\ref{sec:Appendix_C}).

\begin{figure}[htb]
\centering
\includegraphics[width=0.49\textwidth]{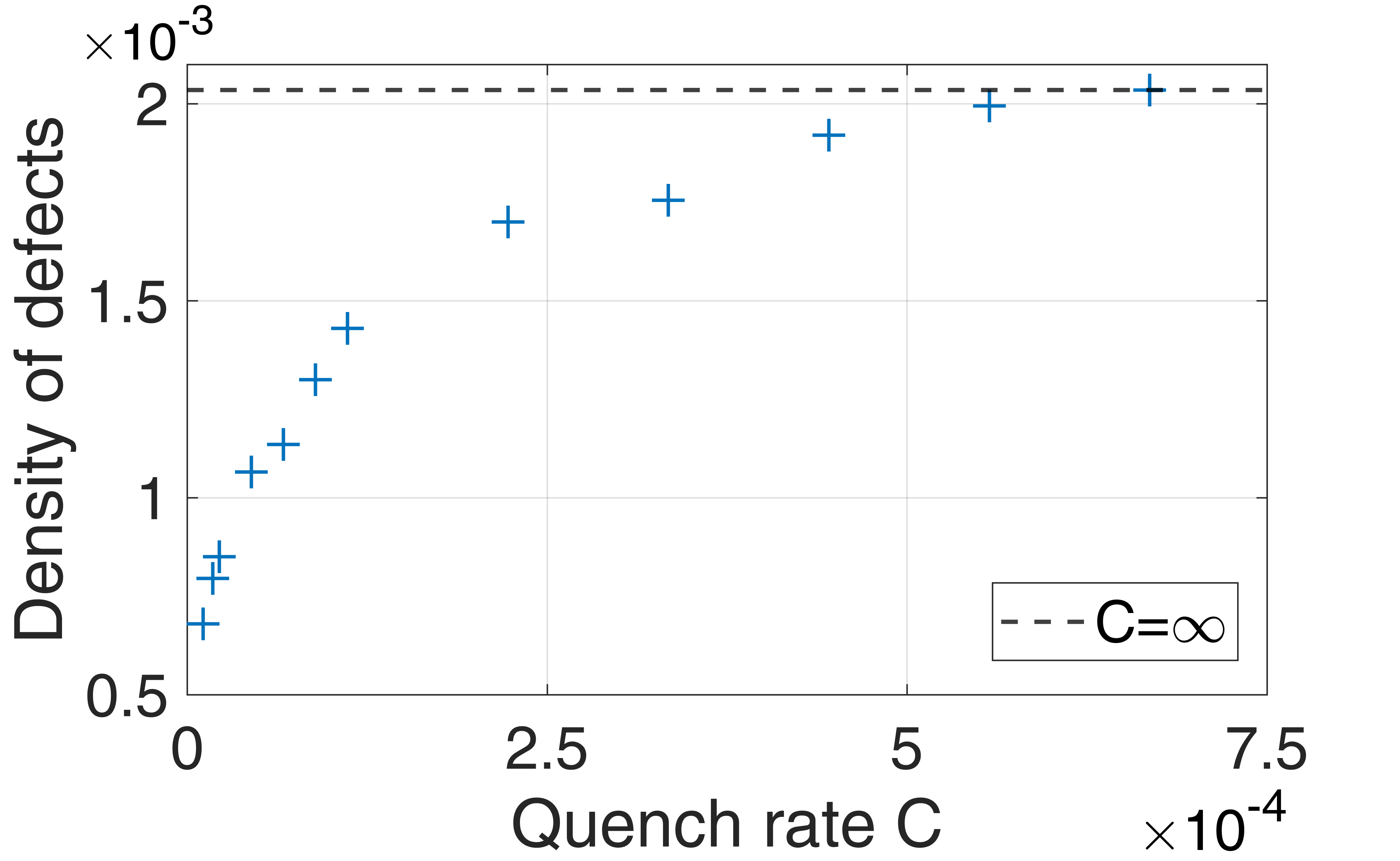}
\caption{Density of defects in the stable state ($T=35000$) as a function of quench rate for a square array of $40000$ oscillators. The results were obtained by averaging over six different realizations. For each realization, the initial conditions of phases was chosen in the range of $\left[-\pi ~\mathrm{to} ~\pi\right]$ and the detuning in the range of $[0 ~\mathrm{to} ~\Omega_{max}\approx\tau_{p}^{-1}\pi/8]$.}
\label{fig:5_gph_of_gph}
\end{figure}

We also determined the density of defects at the stable state as a function of the quench rate, where the density of defects is defined as the number of topological defects divided by the total number of oscillators. The results are presented in Fig.~\ref{fig:5_gph_of_gph}. As evident, the density of defects at the stable state saturates beyond $C\approx5\times 10^{-4}$. As the quench rate decreases, the density of defects is reduced. 

\begin{figure}
\centering
\includegraphics[width=0.49\textwidth]{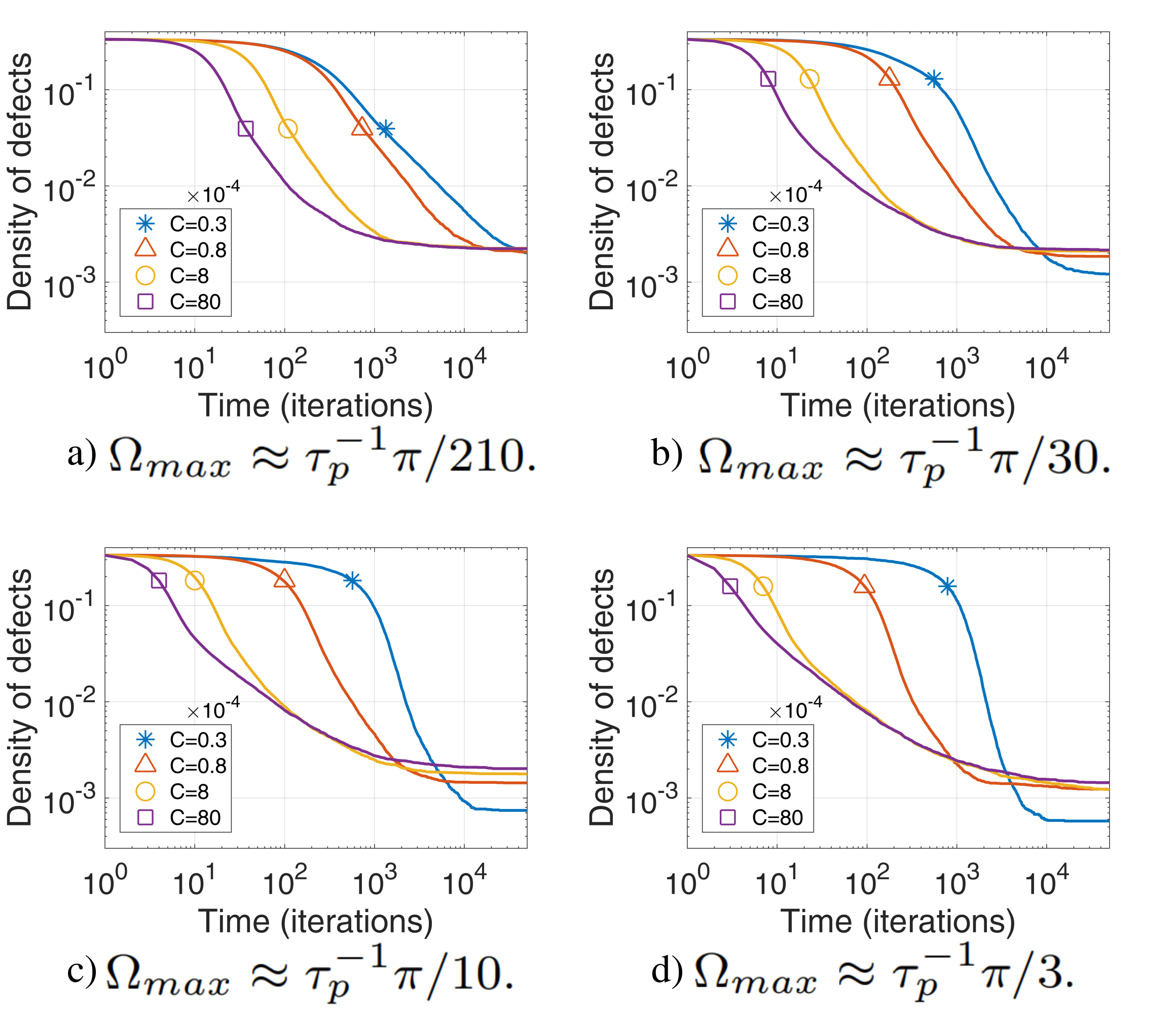}
\caption{Density of defects as a function of time for a square array of $40000$ oscillators, for different coupling rates $C$ and detuning spread $\Omega_{max}$. The inversion of two time scales in the slow decay regime of defects becomes significant as the detuning becomes stronger.  The results were obtained by averaging over six different realizations. For each realization, the initial conditions of phases was chosen in the range of $\left[-\pi ~\mathrm{to} ~\pi\right]$ and the detuning in the range of $[0 ~\mathrm{to} ~\Omega_{max}]$.}
\label{fig:6_ramp_diff_f_defect}
\end{figure}

Next, we calculated the density of defects as a function of time for different maximal detuning $\Omega_{max}$. The results are presented in Fig.~\ref{fig:6_ramp_diff_f_defect}. As evident, when $\Omega_{max}$ is very small, the system converges to the same number of topological defects for all quench rates. As $\Omega_{max}$ increases, there is more inversion of the two competing time scales and less topological defects at long times, indicating an increase of the ordering of the system in stable state. Such effect was demonstrated in a disorder-induced ordering phenomena \cite{anderson,fisher,guillamon}.

\section{Time-dependent detuning}
In this section, we exchange the role of the coupling strength and the detuning between the oscillators in the KZM. Specifically, we use the maximal detuning $\Omega_{max}$ as the control parameter (linearly varied in time) and let the coupling strength between the oscillators be constant in time. Specifically:
\begin{equation}
\Omega_{max}(T)=
\begin{cases}
\left[1-CT\right]\tau_{p}\Omega_{max}, & \text{if}\ \left[1-CT\right]\geq 0 \\
0, & \text{otherwise.} 
\end{cases}
\label{eq:4_ramping_O}
\end{equation}

\begin{figure}
\centering
\includegraphics[width=0.49\textwidth]{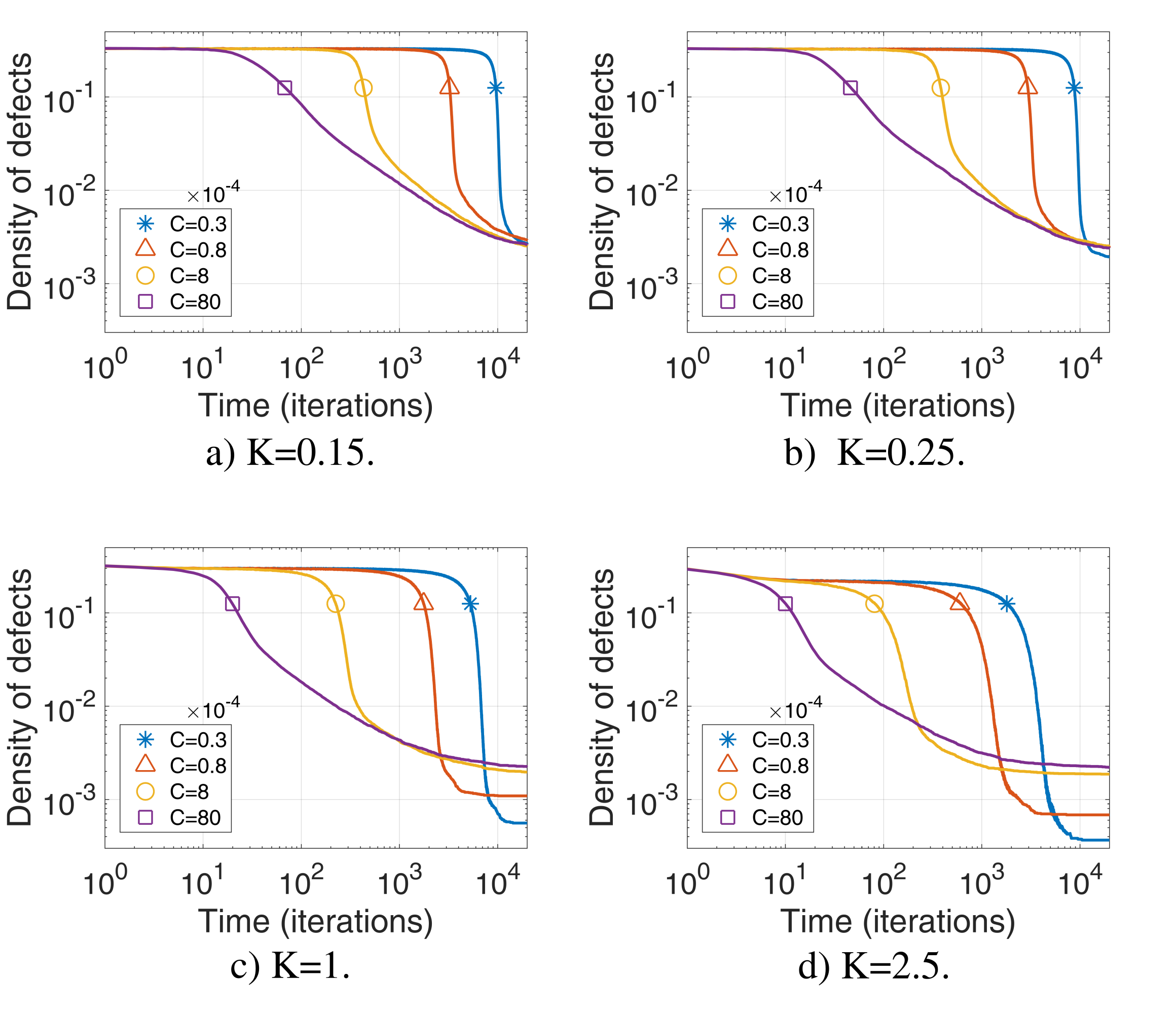}
\caption{Density of defects as a function of time for different quench rates of the detuning spread and coupling strengths for a square array of $40000$ oscillators. The results were obtained by averaging over six different realizations. }
\label{fig:7_ramp_diff_K_defect}
\end{figure}

After incorporating Eq.~(\ref{eq:4_ramping_O}) into Eq.~(\ref{eq:1_kuramoto_model}), we calculated the density of defects as a function of time for different quench rates of $\Omega_{max}$ and (constant) coupling strengths $K$. The results are presented in Fig.~\ref{fig:7_ramp_diff_K_defect}. As evident, when $K$ is small, the system converges to the same number of topological defects for all quench rates. As $K$ increases, there is more inversion of the two competing time scales and less topological defects at long times, indicating an increase of the ordering of the system in stable state. Comparing the results of Fig.~\ref{fig:6_ramp_diff_f_defect} and Fig.~\ref{fig:7_ramp_diff_K_defect} confirm that quenching the coupling strength in a fixed spread of detuning is equivalent to quenching the spread of detuning in a fixed coupling strength. This reflects the close relation between maximal detuning and coupling strength in the Kuramoto model. 
\section{Dissipative topological defects in a one-dimensional ring}
In this section, we consider a one-dimensional ring array geometry of $M=50$ coupled oscillators, i.e. a chain with periodic boundary conditions. The detuning between the oscillators is randomly initialized in the range of 
$[0 ~\mathrm{to} ~\Omega_{max}\approx\tau_{p}^{-1}\frac{\pi}{34}]$. The coupling strength between the oscillators varies linearly in time, Eq.~(\ref{eq:3_ramping_K}), and the coupling occurs over nearest neighbors oscillators only.

\begin{figure}
\centering
\includegraphics[width=0.45\textwidth]{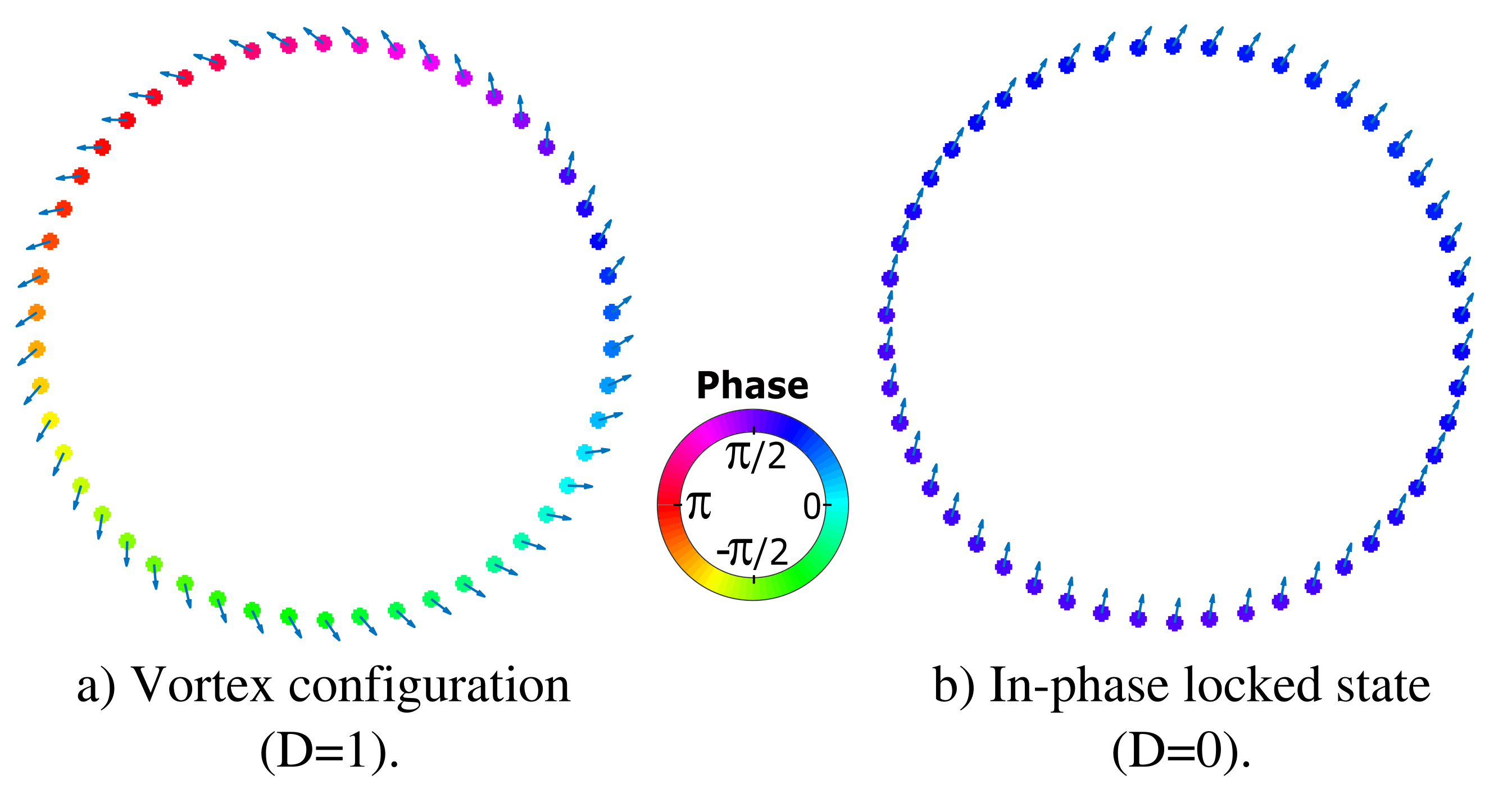}
\caption{Two examples of the phases of $50$ oscillators in a ring array at the stable state ($T=150000$), as obtained from two different realizations. For each realization, the initial conditions of phases was chosen in the range of $\left[-\pi ~\mathrm{to} ~\pi\right]$ and the detuning in the range of $[0 ~\mathrm{to} ~\Omega_{max}\approx\tau_{p}^{-1}\frac{\pi}{34}]$. The coupling was varied linearly in time with a quench rate of $C=3\times 10^{-6}$.}
\label{fig:8_ring_defect}
\end{figure}

For a ring array geometry, the defect number $D$ is globally defined for the whole array as \cite{vishwa, Palopticsexpress}:
\begin{equation}
D=\frac{1}{2\pi}\sum_{m=1}^{M}\left\{\phi_{m+1}-\phi_{m}\right\},
\label{eq:5_defect_number_1D}
\end{equation} 
where $\{ \}$ wrap angle in $\left[-\pi ~\mathrm{to} ~\pi\right]$. $D>0$ corresponds to a vortex and $D<0$ to an anti-vortex. After incorporating Eq.~(\ref{eq:3_ramping_K}) into Eq.~(\ref{eq:1_kuramoto_model}), we calculated the phases of the $50$ oscillators at the stable state ($T=150000$) for two different realizations, each with a different initial phase distribution, and both with a quench rate of $C=3\times 10^{-6}$. The results are presented in Fig.~\ref{fig:8_ring_defect}. Two different stable states are observed: a vortex in Fig.~\ref{fig:8_ring_defect}(a) with $D=1$ and $r\approx0$; and an in-phase-locked state in Fig.~\ref{fig:8_ring_defect}(b) with $D=0$ and $r\approx1$. 

\begin{figure}[htb]
\centering
\includegraphics[width=0.4\textwidth]{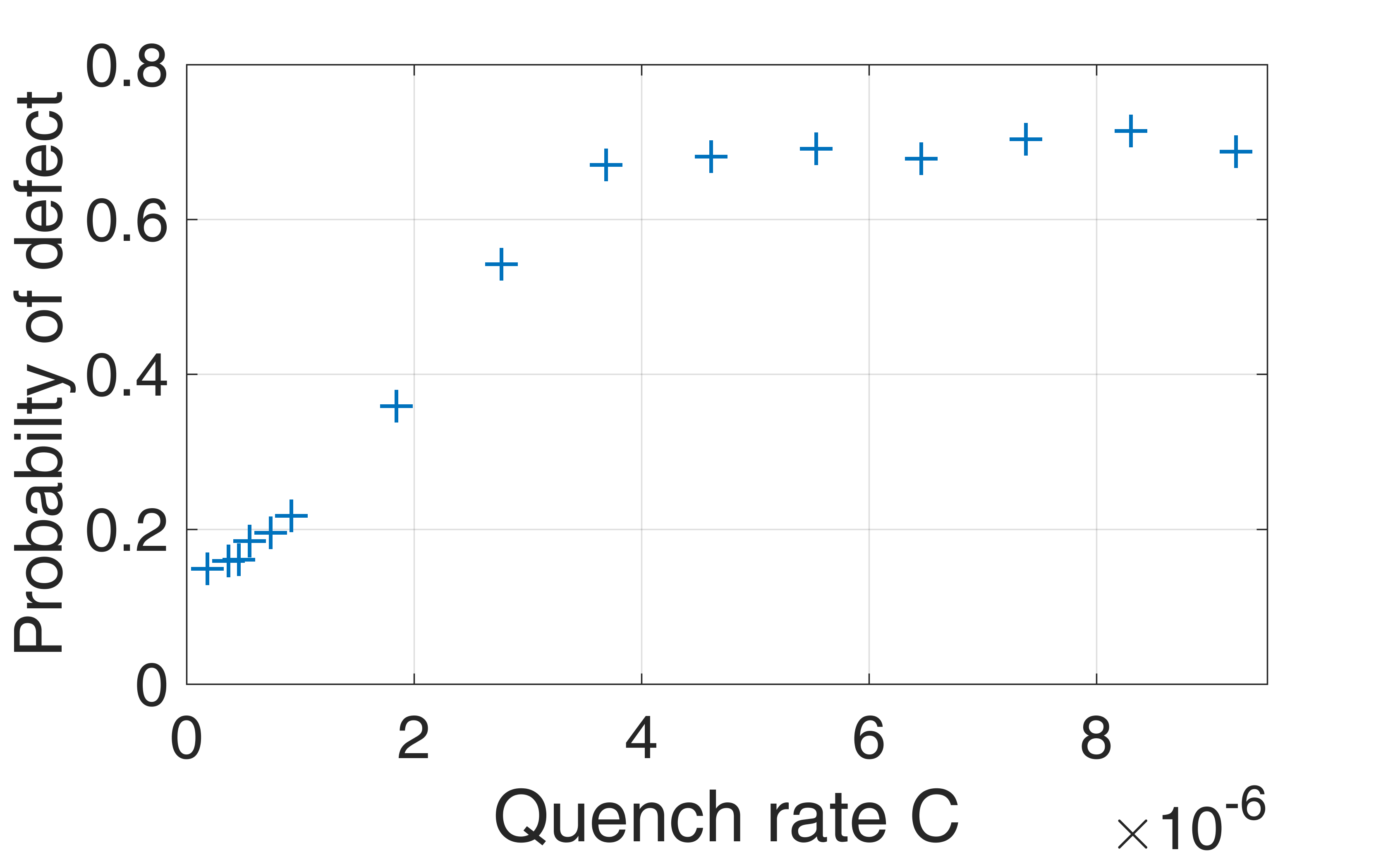}
\caption{Probability of having topological defects as a function of quench rate for a ring of 50 oscillators. For each quench rate, the probability of defects was determined from $1500$ realizations. For each realization, the initial conditions of phases was chosen in the range of $\left[-\pi ~\mathrm{to} ~\pi\right]$ and the detuning in the range of $[0 ~\mathrm{to} ~\Omega_{max}\approx\tau_{p}^{-1}\frac{\pi}{34}]$.}
\label{fig:9_defect_prob_t_1D}
\end{figure}

The statistics on the number of topological defects is obtained by averaging over $1500$ realizations. The probability of having a topological defect is the ratio of the number of realizations that lead to a topological defect at the stable state over the total number of realizations. The results are presented in Fig.~\ref{fig:9_defect_prob_t_1D}. As evident, the probability of having topological defects at the stable state saturates to about $70\%$ at quench rates $C=4\times 10^{-6}$ and above. For lower quench rates, the probability of having topological defects monotonically decreases, in analogy to KZM in one dimension \cite{vishwa}. However, as the quench rate goes to zero, the probability of having a topological defect approaches a finite non-zero value ($\approx0.15$ for our system). This value grows up for lower values of $\Omega_{max}$ indicating the need of finite detuning spread to suppress topological defects for slow quench rates.

\section{Concluding Remarks}
We numerically investigated the dynamics of dissipative topological defects in one and two-dimensional arrays of coupled oscillators, by using the Kuramoto model. An analogy with the Kibble-Zurek mechanism was shown, where the dynamics of the topological defects was governed by two competing time scales and their density scaled with the coupling rate. In the short time scale regime, the number of topological defects rapidly decreased in time with increasing coupling strength. However, in the stable state, the number of topological defects reduced with the coupling rate. This reduction depends both on the detuning between the oscillators and coupling rate. For a certain coupling rate, introducing detuning between the oscillators initially in the system lead to a more ordered stable state with less topological defects (disorder-induced ordering). 

In the Kibble-Zurek mechanism, the number of topological defects obeys a power law as a function of the quench rate. This power law was related to the critical exponents associated to the correlation length and the relaxation time of the system. Accordingly, we plan to extend our investigations to determine whether the scaling of our system will exhibit any power-law behavior, and so as directly resemble the Kibble-Zurek mechanism. Then, we could analyze the critical exponents of scaling behavior, and find the universality class of coupled phase oscillators with dissipative topological defects.
\begin{acknowledgments}
We wish to acknowledge the Israel Science Foundation and the Israel-U.S. Binational Science foundation for their support.
\end{acknowledgments}
\appendix
\section{Defect number D for a square array and time evolution of coupling and detuning.}
\label{sec:Appendix_A}
For a square array of coupled oscillators, the defect number $D_{ij}$ is defined for each oscillator of the array, as:
\begin{eqnarray}
D_{_{ij}}=&\frac{1}{2\pi}\left[\{\phi_{_{i(j+1)}}-\phi_{_{ij}}\}+\{\phi_{_{(i+1)(j+1)}}-\phi_{_{i(j+1)}}\}\right.\nonumber\\&\left.+\{\phi_{_{(i+1)j}}-\phi_{_{(i+1)(j+1)}}\}+\{\phi_{_{ij}}-\phi_{_{(i+1)j}}\}\right],
\label{eq:S1_defect_number_2D_square} 
\end{eqnarray} 
where the brackets $\{ \}$ wrap phases in the range of $\left[-\pi ~\mathrm{to} ~\pi\right]$, $i$ the row number and $j$ the column number ($i$ and $j$ in the range of $\left[1 ~\mathrm{to} ~M\right]$). For a ring array of coupled oscillators, the defect number $D$ is defined by Eq.~\eqref{eq:5_defect_number_1D} for the whole array.\\

\begin{figure}[htbp]
\centering
\includegraphics[width=0.49\textwidth]{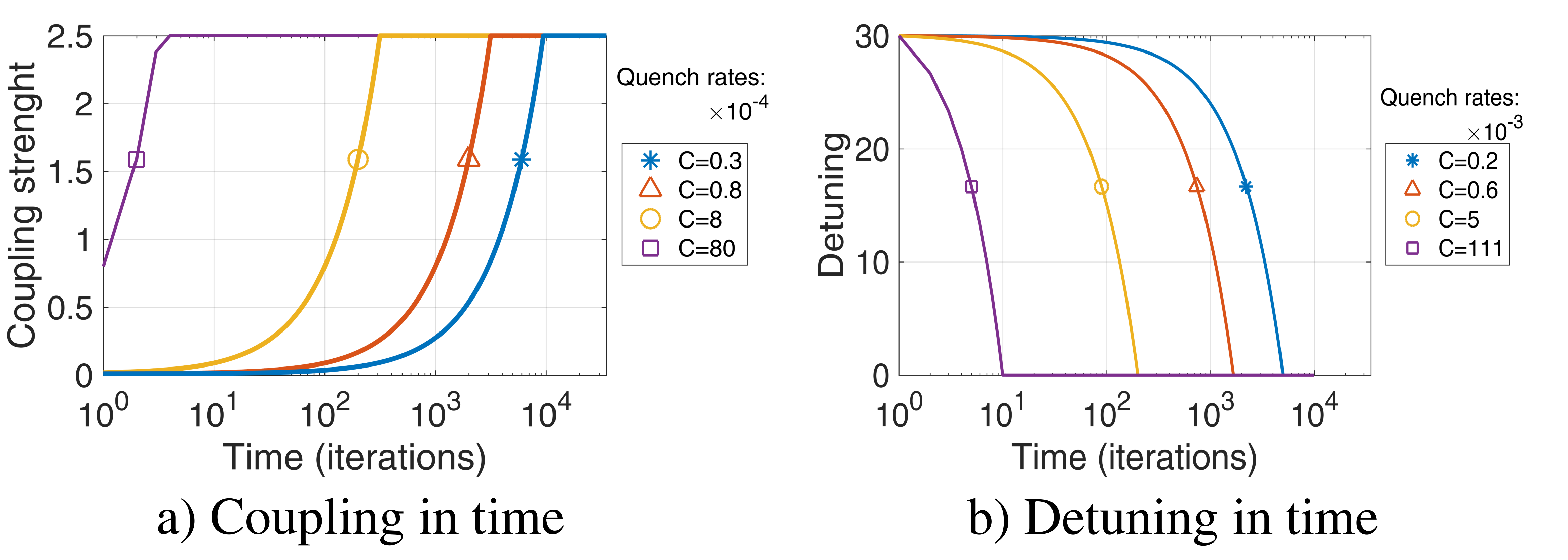}
\caption{Variation in time of the coupling strength and detuning for different quench rates. The quench rate determines the speed of the linear variation of the coupling strength. Note the logarithmic scale along the horizontal axis.}
\label{fig:S1_ramping_K_O}
\end{figure}

We calculated the coupling strength and detuning as a function in time using Eq.~(\ref{eq:3_ramping_K}) and Eq.~(\ref{eq:4_ramping_O}). The results are presented in Fig.~\ref{fig:S1_ramping_K_O}. Figure \ref{fig:S1_ramping_K_O}(a) shows the variation of coupling in time for the different quench rates that were used to calculate the results of Figs.~\ref{fig:4_effect_eg_1},~\ref{fig:5_gph_of_gph} and ~\ref{fig:6_ramp_diff_f_defect}. Figure ~\ref{fig:S1_ramping_K_O}(b) shows the variation of detuning in time for the quench rate that was used to calculate the results of Fig.~\ref{fig:7_ramp_diff_K_defect}.
\section{Dissipation of topological defects as a function of array size}
\label{sec:Appendix_B}
\begin{figure}[htbp]
\centering
\includegraphics[width=0.49\textwidth]{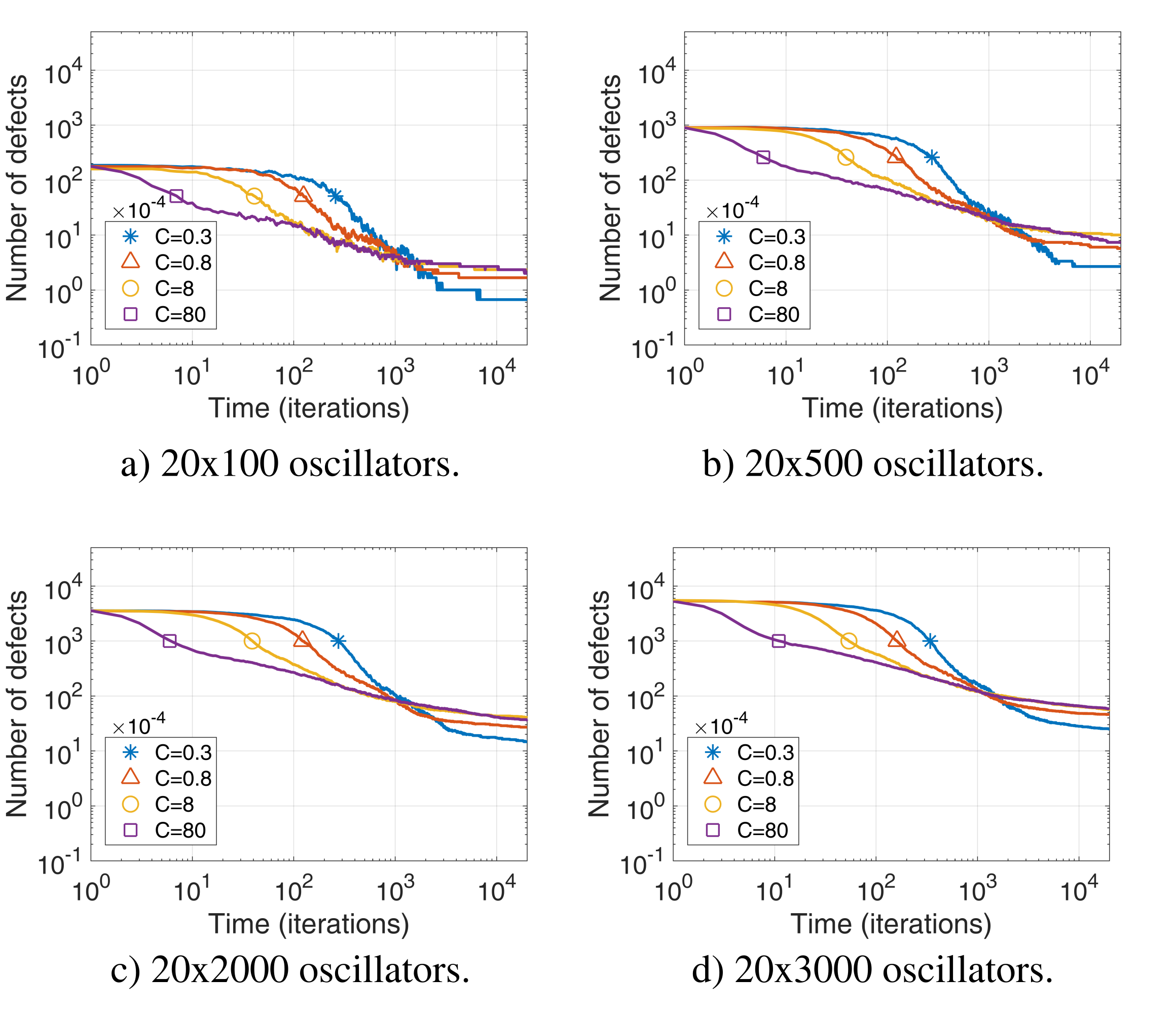}
\caption{The number of topological defects as a function of time for different coupling rates and for different number of oscillators. As evident, the number of oscillators in the system does not affect the inversion of the two time scales, provided that the system size (number of oscillators) is not too small. The results were obtained by averaging over six different realizations.}
\label{fig:S2_KZM_vs_size_system}
\end{figure}

We calculated the number of topological defects as a function of time for four different quench rates and for different sizes of the system (number of oscillators).  Aside from the system size, all other parameters were the same as those used for obtaining the results of Fig.~\ref{fig:4_effect_eg_1}. The results are presented in Fig.~\ref{fig:S2_KZM_vs_size_system}, where the number of oscillators was relatively large and varied from $2000$ to $60000$. As depicted in Fig.~\ref{fig:S2_KZM_vs_size_system}, the smaller the system size, the smaller the number of topological defects. Yet, the inversion of the two competing time scales occurs at the same time for the four different system sizes. Surprisingly, the system size does not govern the inversion of the two time scales.

\section{Distribution of the topological defects at T=1}
\label{sec:Appendix_C}
\begin{figure}[htb]
\centering
\includegraphics[width=0.49\textwidth]{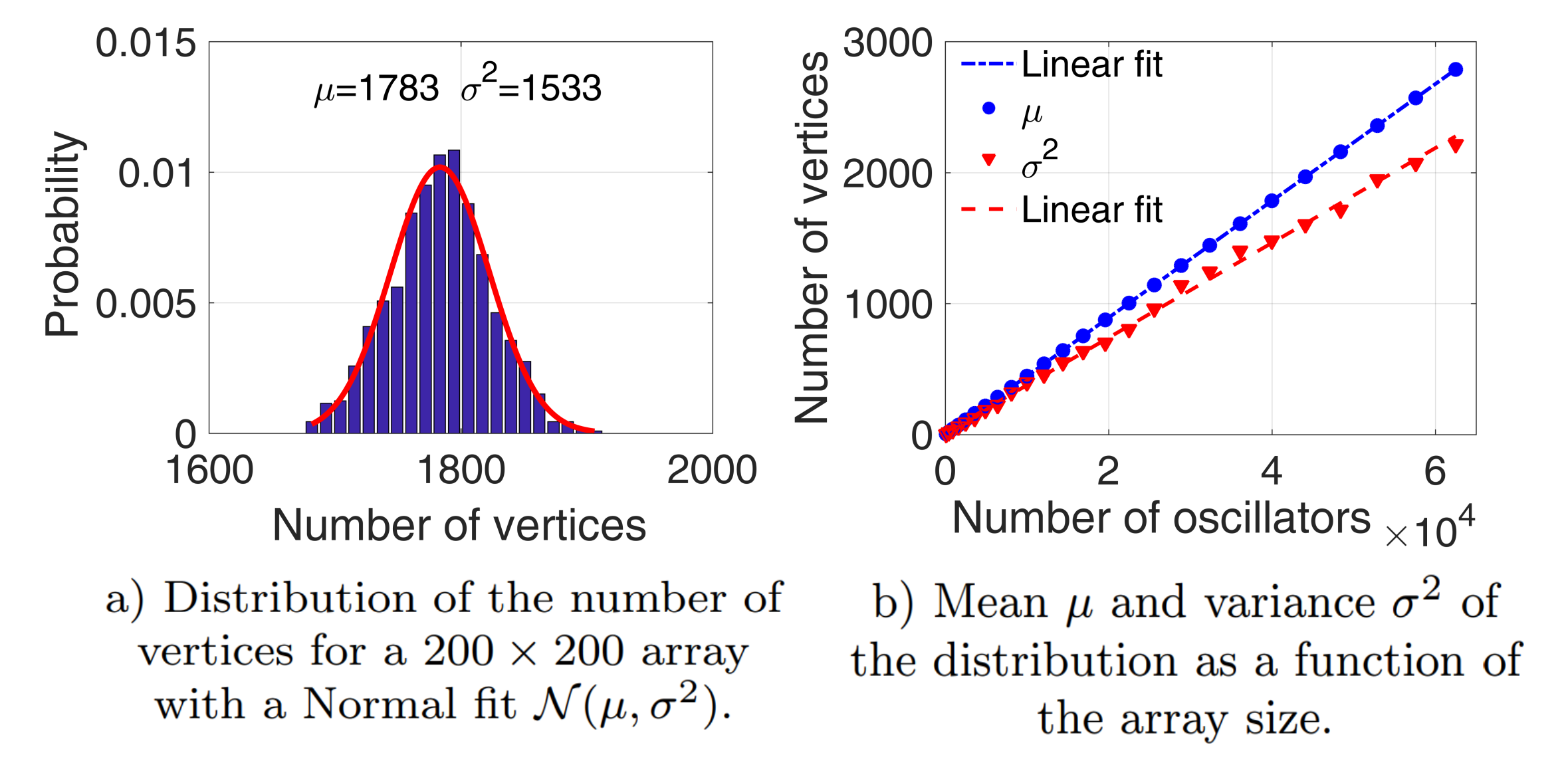}
\caption{Probability distribution of the number of vortices at the initial state and the mean and variance of the Normal fit as a function of the number of oscillators. a) Probability distribution of the number of vortices at initial state ($T=1$) for a square array of $200\times200$ oscillators, using $10000$ realizations.  b) Mean and variance of the Normal fit as a function of the numbers of oscillators}
\label{fig:S3_distribution_T_0}
\end{figure}

We determined the probability distribution of the number of number of vortices, anti-vortices and topological defects at the initial state (at $T=1$). Since the definition of an anti-vortex is the opposite of a vortex, the probability distribution of the number of vortices must be identical to that of the number of anti-vortices, and proportional to the distribution of the number of topological defects. For this reason, only the distribution of the number of vortices at the initial state was considered. The probability distribution was calculated using $10000$ realizations, where in each realization the number of vortices was calculated for a square array of $M$ oscillators with random phase initialized in the range $\left[-\pi ~\mathrm{to} ~\pi\right]$.

We calculated the probability distribution of the number of vortices at the initial state for a square array of $40000$ oscillators and the mean and variance of the Normal fit as a function of the number of oscillators. The results are presented in Fig.~\ref{fig:S3_distribution_T_0}. Figure ~\ref{fig:S3_distribution_T_0}(a) shows the probability distribution as a function of the number of vortices at the initial state. As evident by the red curve, the probability distribution fit well the Normal distribution. Using the Normal fit, we calculated the corresponding mean $\mu$ and variance $\sigma^{2}$ as a function of the number of oscillators. The results are shown in Fig.~\ref{fig:S3_distribution_T_0}(b). The mean and the variance of the Normal fit are a linear function of the number of oscillators $M$ \citep{distrib}. 

\bibliography{apssamp}
\end{document}